\documentclass[]{aa}
\usepackage{graphicx}
\usepackage{natbib}
\bibpunct{(}{)}{;}{a}{}{,}

\usepackage{txfonts}
\begin{document}
   \title{Detection of abundant solid methanol toward young low mass stars \thanks{Based on observations
obtained at the European Southern Observatory, Paranal, Chile, within
the observing program 69.C-0441. }}
   \titlerunning{Abundant solid methanol toward low mass YSOs}
   \author{K. M. Pontoppidan \inst{1} \and E. Dartois\inst{2} \and E. F. van Dishoeck\inst{1} \and W.-F. Thi\inst{3} \and L. d'Hendecourt\inst{2}}

\institute{Leiden Observatory, P.O.Box 9513, 2300 RA Leiden, The Netherlands\and
{Institut d'Astrophysique Spatiale, B{\^a}t. 121, CNRS UMR8617, Universit{\'e} Paris XI, 91405 Orsay Cedex, France} \and
{Astronomical Institute ''Anton Pannekoek'', University of Amsterdam, Kruislaan 403, 1098 SJ Amsterdam, The Netherlands}}

 \offprints{K. M. Pontoppidan \email{pontoppi@strw.leidenuniv.nl}}
 \date{received/accepted}

   \abstract{We present detections of the absorption band at $\rm
3.53~\mu m$ due to solid methanol toward three low-mass young stellar
objects located in the Serpens and Chameleon molecular cloud
complexes. The sources were observed as part of a large spectroscopic survey of
$\approx 40$ protostars. This is the first detection of solid methanol in the
vicinity of low mass ($M\lesssim1~M_{\odot}$) young stars and shows
that the formation of methanol does not depend on the
proximity of massive young stars. The abundances of solid methanol
compared to water ice for the three sources are in the range 15-25\%
which is comparable to those for the most methanol-rich massive sources known.
The presence of abundant methanol in the circumstellar environment of
some low mass young stars has important consequences for the formation
scenarios of methanol and more complex
organic species near young solar-type stars.
}

   \maketitle

\section{Introduction}
The presence and origin of complex organic molecules in protostellar
regions and their possible incorporation in protoplanetary disks is an
active topic of research, both observationally and through laboratory
simulations. Large organic molecules such as CH$_3$OCH$_3$ and
CH$_2$CH$_3$CN have been detected with enhanced abundances in
so-called `hot cores' around massive young stars \citep[e.g.][]{Hatchell98, GibbThesis}. 
In the laboratory, UV photolysis of
ice mixtures of species prepared with interstellar abundances followed by thermal warm-up to 300~K is
known to lead to a wealth of complex species such as carboxylic acids
and esters \citep{Briggs92}, hexamethylene-tetramine \citep{Bernstein95} and prebiotic molecules including amino acids \citep{Guille, Bernstein02}.
In the ice experiments, methanol is thought to be a key molecule in the production of these
complex molecules. Understanding
the methanol content of interstellar ice mantles and its variation in
different circumstellar environments is therefore essential to test
the validy of the basic assumptions that enter the chemical models and
experimental approach.

So far, solid methanol has only been detected along lines of sight
associated with intermediate- to high-mass protostars, with abundances
ranging from 5\% up to 30\% with respect to the dominant ice mantle
component, H$_2$O \citep{Manu99}.  Extensive searches toward low-mass protostars have
resulted in typical upper limits of a few \% \citep{Chiar96,Brooke99}.  In contrast, gas-phase methanol has been found toward both
high- and low-mass YSOs with evidence for abundance jumps of factors
of 100--1000 in the inner part of the envelope, consistent with ice
evaporation \citep[e.g.][]{vdtak00,Fredrik02,Bachiller95}.
Still, the observed gas-phase
column densities of methanol are usually $10-10^4$ times less than those observed in the
solid phase \citep[e.g.][]{Fredrik02}. 
In hot cores, a rapid high-temperature gas-phase
chemistry is triggered by the evaporation of methanol-rich ices, which
forms high abundances of large organic molecules for a period of
$10^4-10^5$ yr \citep[e.g.][]{Charnley92, Rodgers01}. Understanding the origin 
of solid $\rm CH_3OH$ and its variation from source to source is a key ingredient for these models.

As part of a large VLT-ISAAC 3--5 $\mu$m spectral survey toward young
low mass stars we have detected an absorption feature at $\rm 3.53~\mu
m$ towards at least three icy sources out of $\approx 40$, which is attributed to solid
methanol. Two of the sources are part of the dense cluster of young
low-mass stars, \object{SVS 4}, located in the Serpens cloud core \citep{EC89}. The third
source is a more isolated low-mass class I source in the
Chameleon I cloud known as \object{Cha INa 2}  \citep{Persi99}. The spectra presented here 
represent the first detections of solid methanol
toward low mass YSOs. The observations
and analysis are reported in \S 2, whereas the implications
for the chemical evolution of circumstellar matter are discussed in \S 3.

\section{Observations}
\subsection{The $L$-band spectra}
The observations were performed using the Infrared Spectrometer and
Array Camera (ISAAC) mounted on the Very Large Telescope
UT1-Antu of the European Southern Observatory at Cerro Paranal.  $L$-band
spectra were obtained using the low resolution mode and the 0\farcs3
slit resulting in resolving powers of $\lambda/\Delta\lambda=1200$.
\object{SVS 4-5} and \object{SVS 4-9} were observed simultaneously on May 5, 2002. \object{Cha
INa 2} was observed on May 6, 2002.  All spectra were recorded during good weather conditions.  The
data were reduced using our own IDL routines, which are described in
\cite{me03}. Correction for telluric absorption features was done
using the early-type standard stars \object{BS 4773} (B5V) and \object{BS 7348} (B8V)
for \object{Cha INa 2} and the \object{SVS 4} sources, respectively. The wavelength
calibration was obtained using the telluric absorption features with
an estimated accuracy of $\rm 0.001~\mu m$. The spectra were flux
calibrated using the standard stars with an estimated uncertainty of
15\%, mainly due to uncertainties in slit losses.
\begin{figure}
\centering
\includegraphics[width=5.5cm,angle=90]{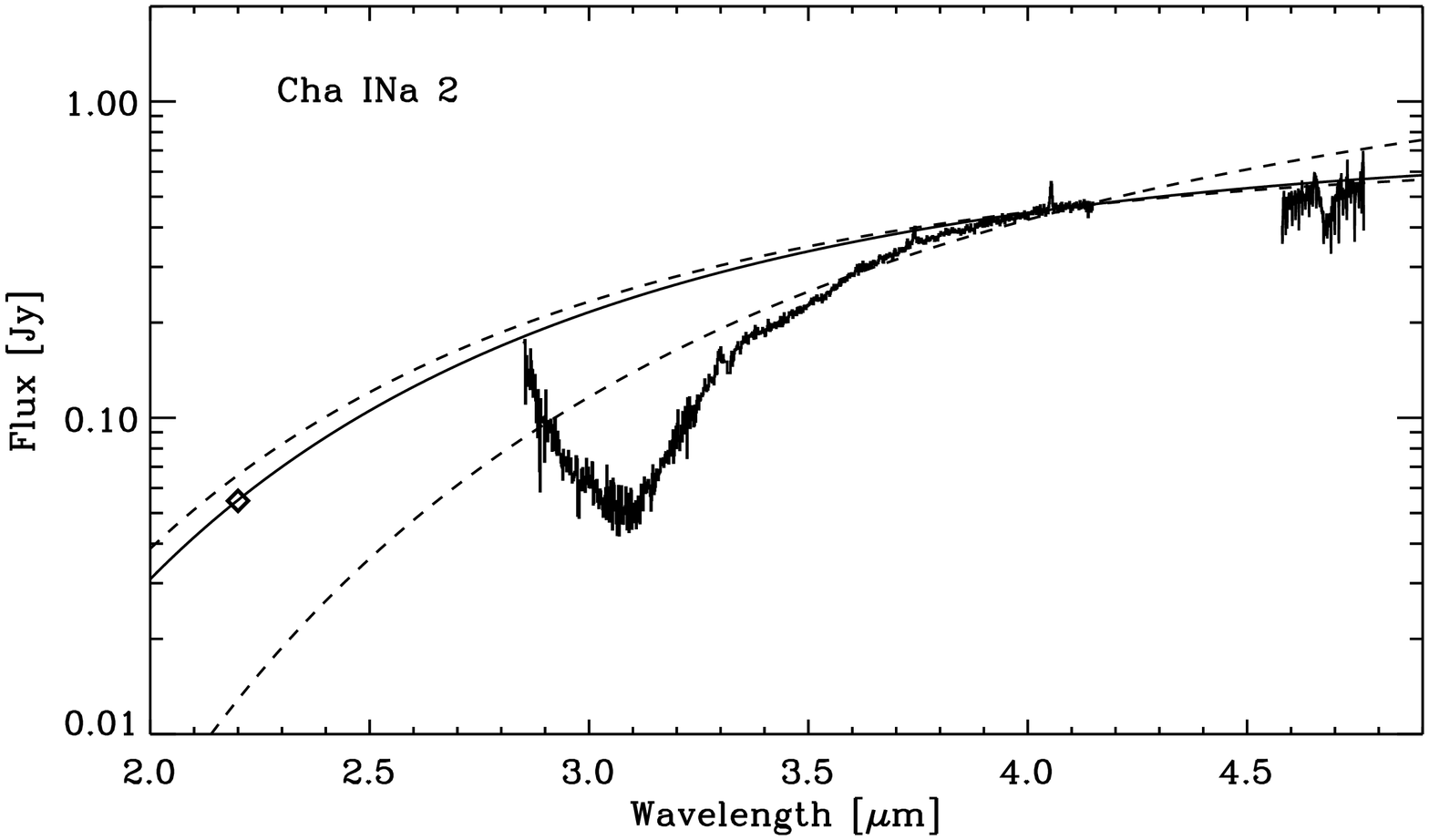}
\includegraphics[width=5.5cm,angle=90]{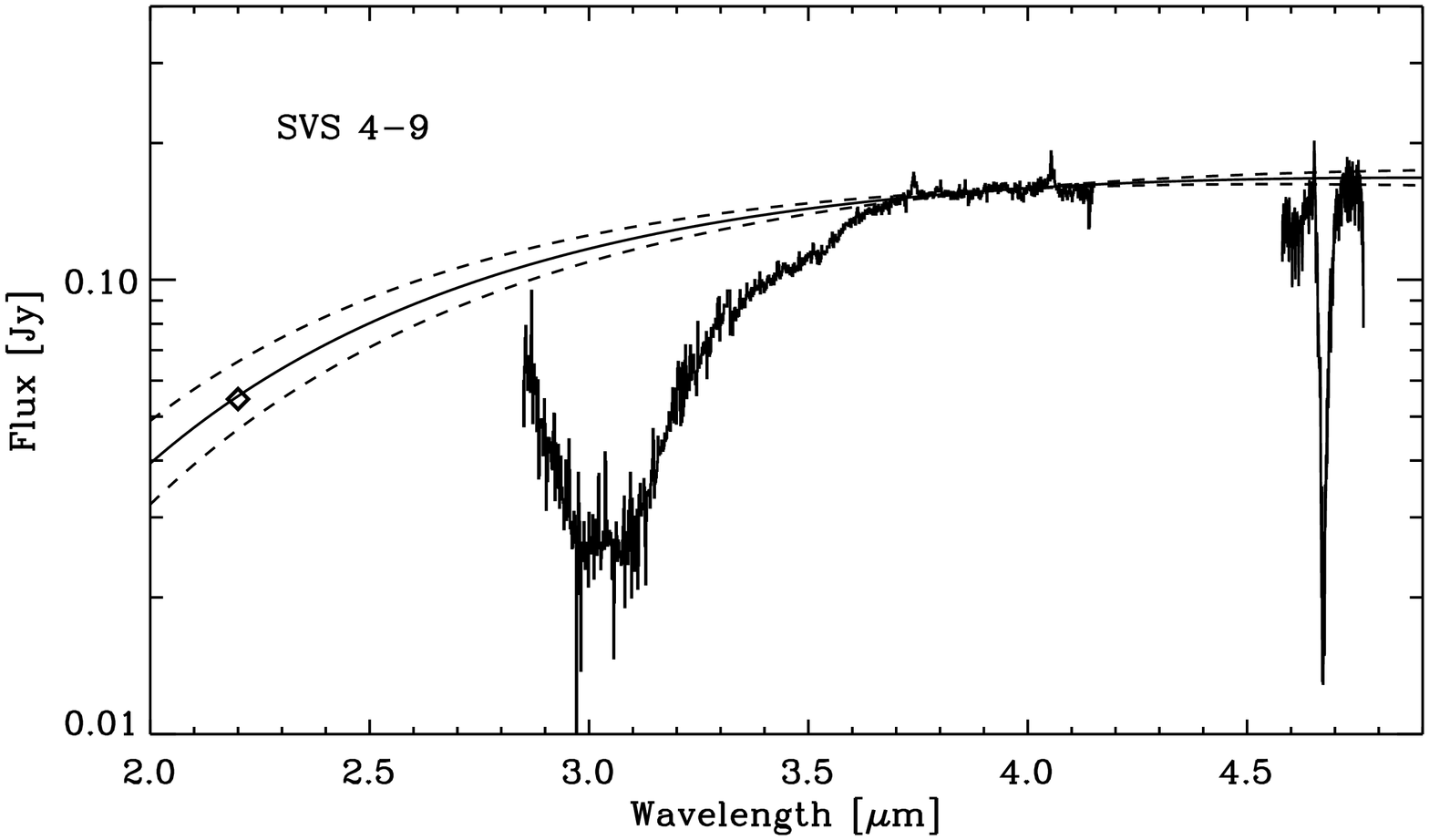}
\includegraphics[width=5.5cm,angle=90]{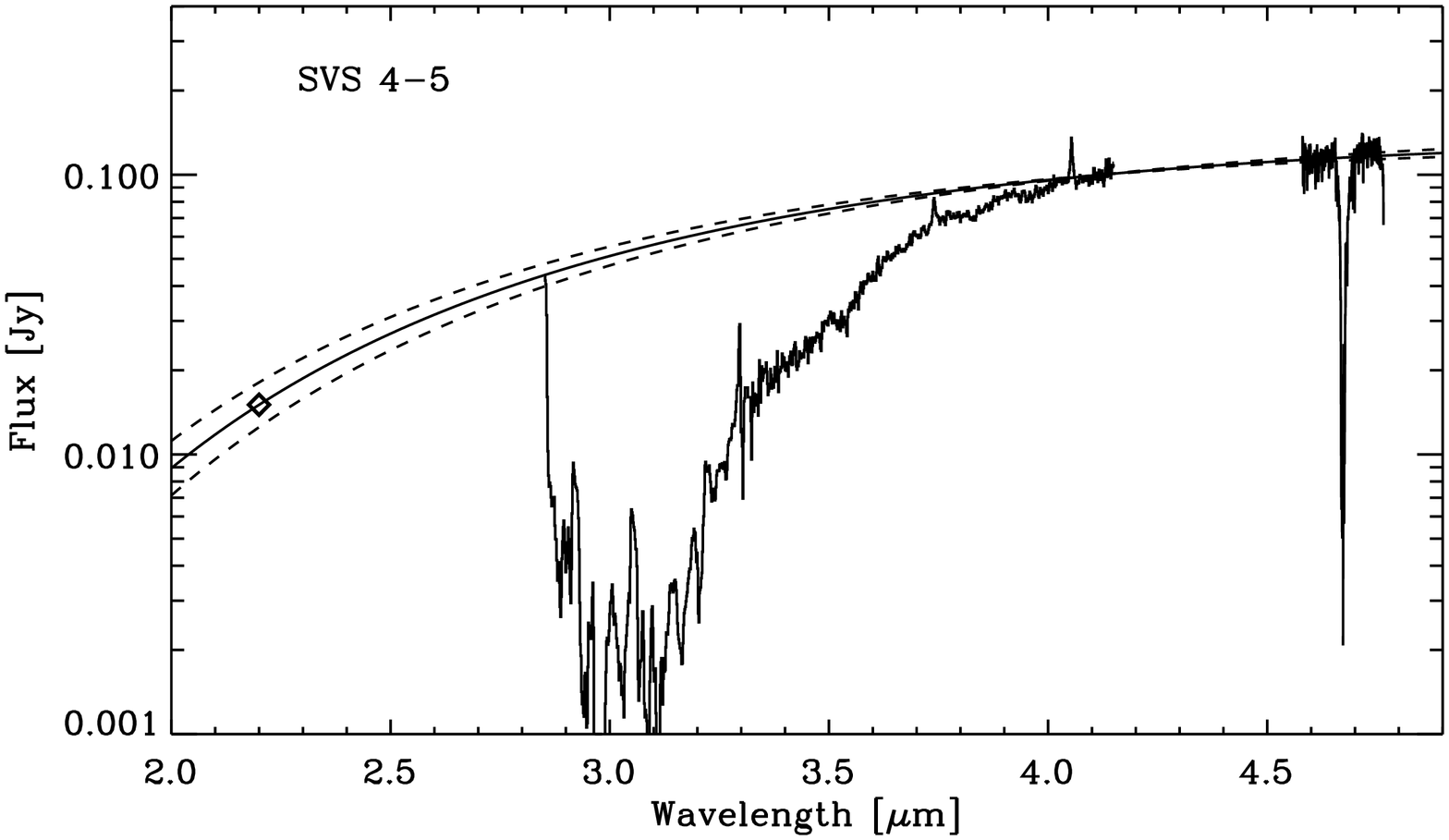}
\caption{Adopted blackbody continua for estimating water band optical
depths (solid curves).  The $K$-band magnitudes (diamonds) for the Serpens sources are taken
from \cite{Giovannetti98}. The adopted $K$-band magnitude for Cha INa
2 is discussed in the text. Dashed curves indicate continua calculated
using a conservative estimate for the uncertainties on the (sometimes
variable) $K$-band fluxes. The spectrum of SVS 4-5 has been smoothed to a resolution of $R=120$
between 2.8 and $\rm 3.3~\mu m$.}
\label{continuaAll}
\end{figure}

\begin{figure*}
\centering
\includegraphics[width=3.5cm,angle=90]{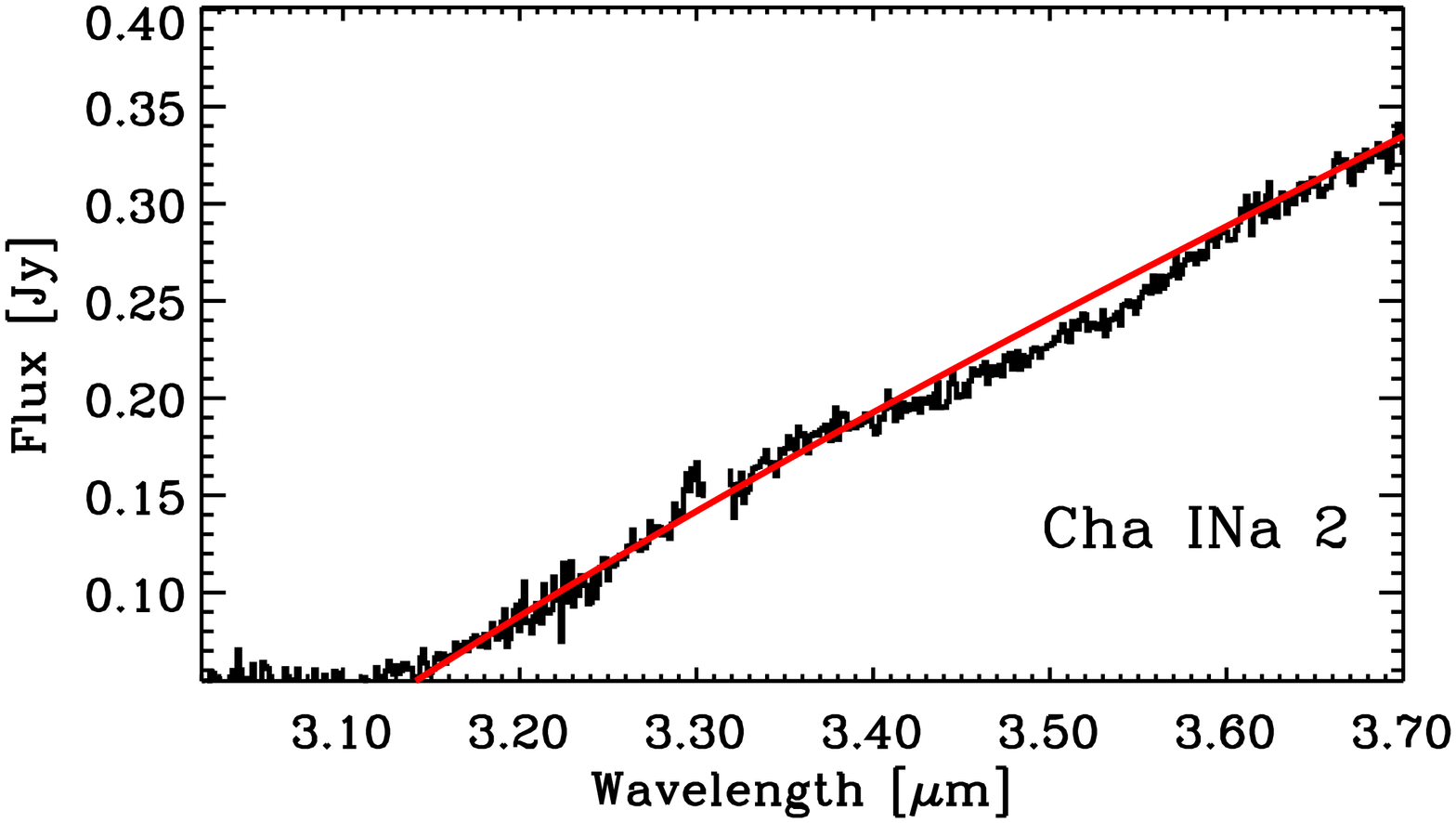}
\includegraphics[width=3.5cm,angle=90]{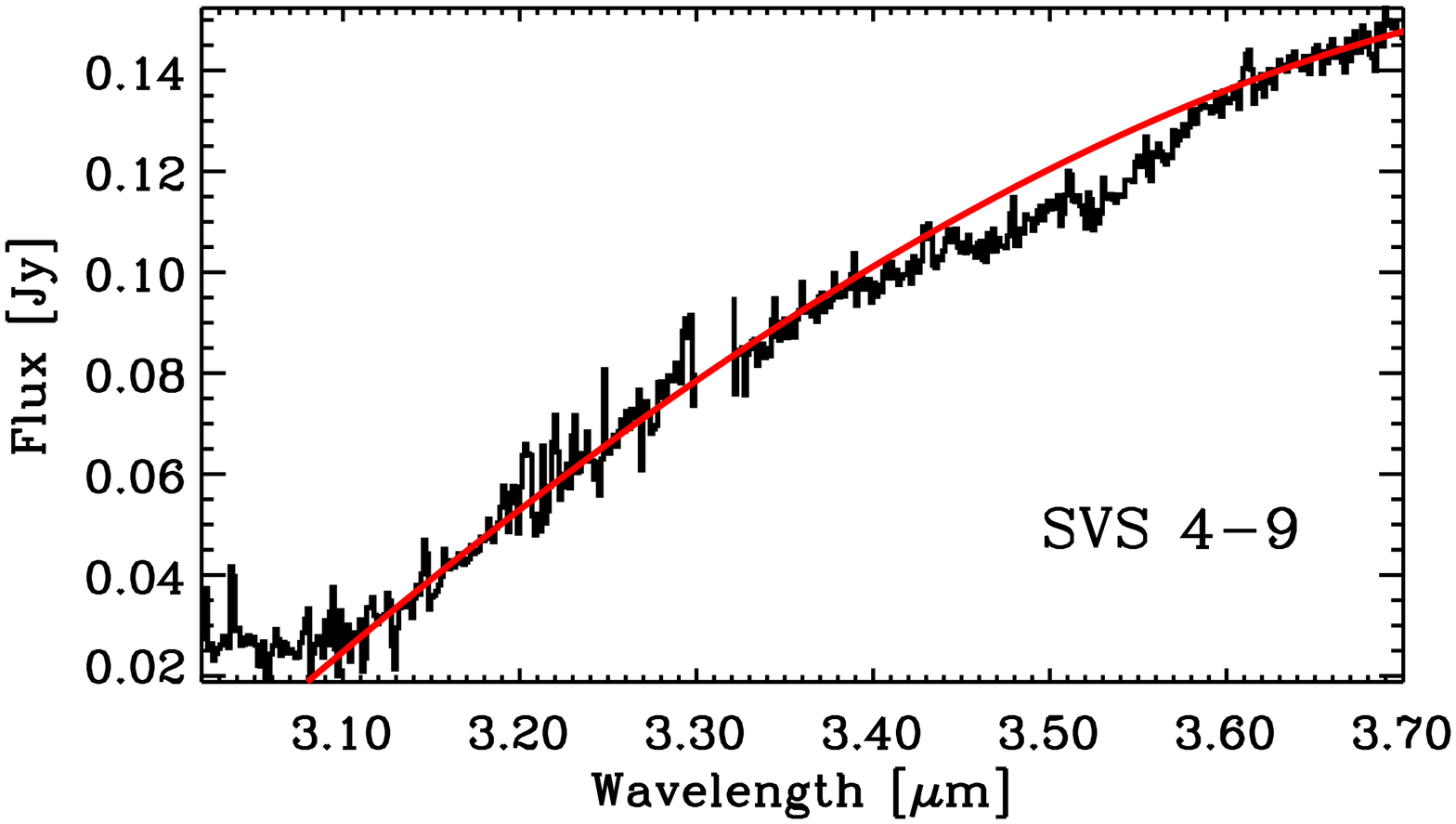}
\includegraphics[width=3.5cm,angle=90]{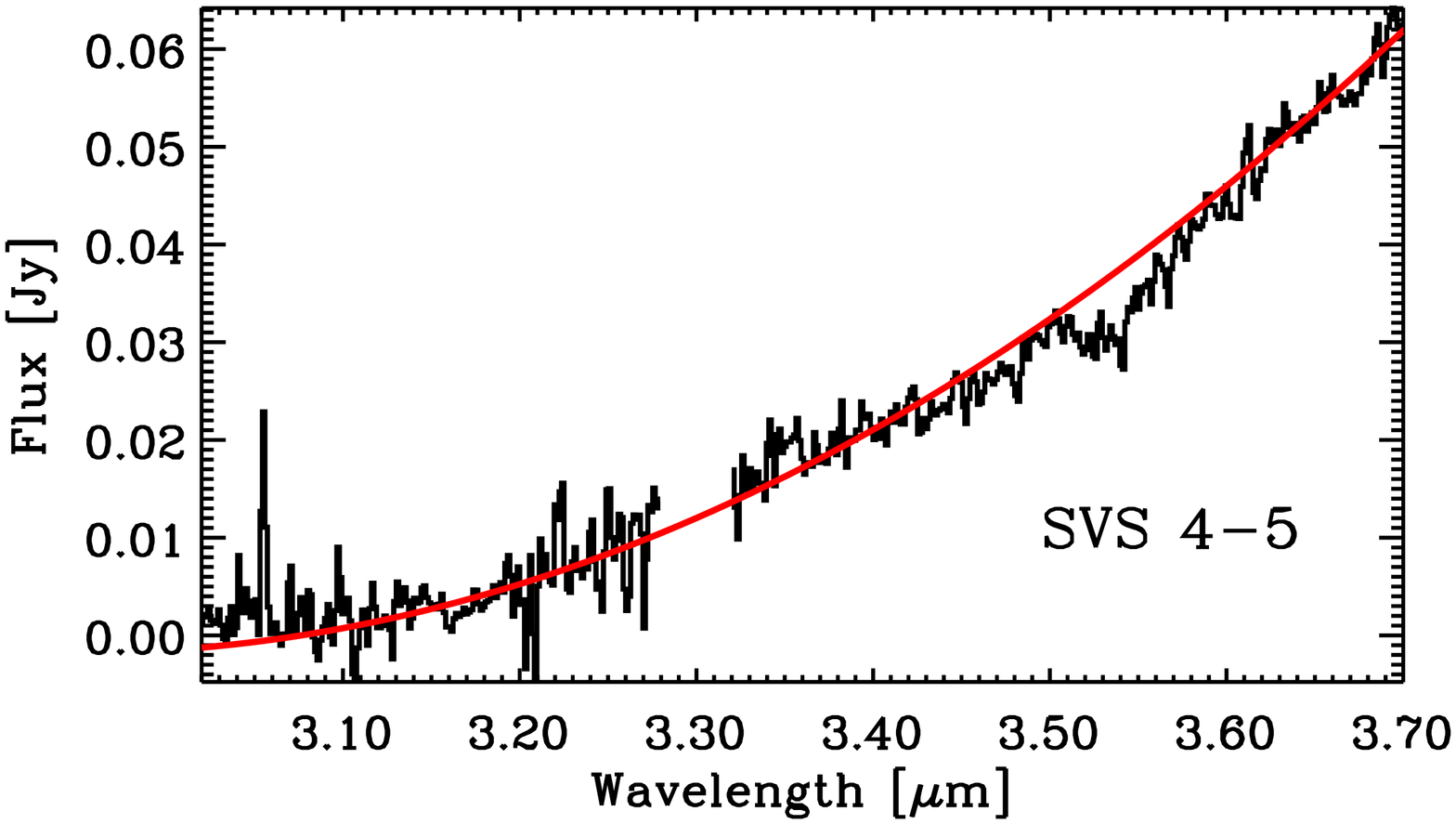}
\caption{Local third order polynomial continua for extracting the 
methanol band profiles. }
\label{continua}
\end{figure*}

\subsection{Derivation of methanol abundances}

The water band profile was obtained by adopting a blackbody continuum
fitted to a $K$-band photometric point from the literature and the
observed $L$-band spectrum from 4.0 to $\rm 4.15~\mu m$ as well as the
$M$-band spectra, which are discussed in detail in \cite{me03}. 
The $\rm 2.8-4.8~\mu m$ spectra are shown in Fig. \ref{continuaAll}. 
The accuracy of the $K$-band points are crucial for estimating the
column density of the water ice bands. However, the $K$-band fluxes have
considerable uncertainties since many young stellar
objects are known to be highly variable and the sources may have
changed their flux level since the photometric
measurement. Indeed, Cha INa 2 is known to vary significantly in the
$K$-band \citep{Carpenter02} and has apparently continuously
brightened with 1.7 mag between March 1996 and April 2000
\citep{Persi99, Kenyon01,Carpenter02} to $K=11.044$. It is
found that at least $K=10.2$ is necessary to fit the continuum defined
by the ISAAC $L$ and $M$-band spectra, consistent with the
assumption that the source has continued to brighten until May 2002
with the same rate. 
The continua used for deriving the optical depths of the water bands
are shown in Fig. \ref{continuaAll} along with continua fitted using
extreme $K$-band fluxes. For the Serpens sources a conservative
estimate of the $K$-band variability is 0.2 mag when comparing with
the photometry of \cite{EC92}. For Cha INa 2 continua using $K=10$ and
$K=12.75$ are also shown.

\begin{figure}
\centering
\includegraphics[width=9cm]{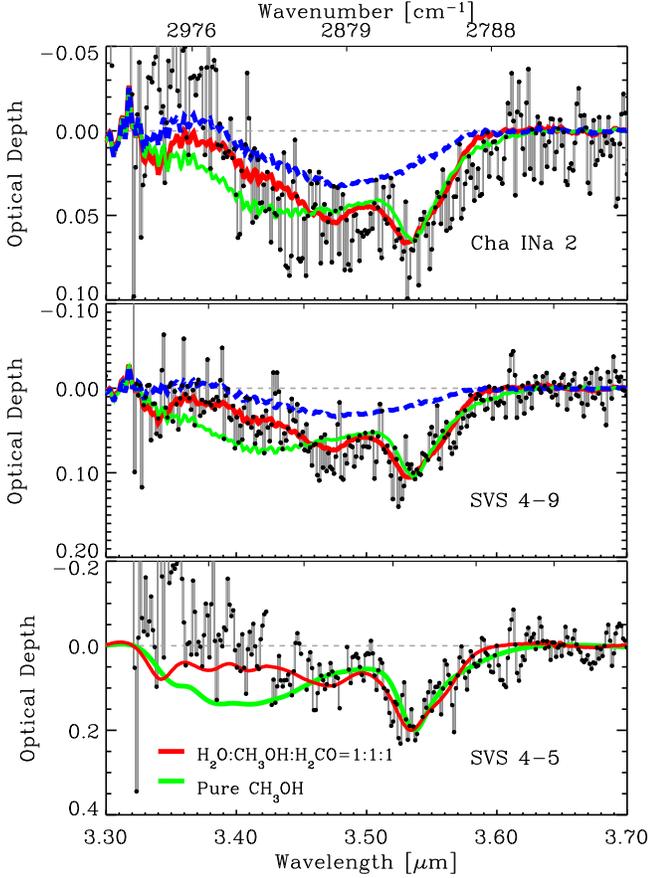}
\caption{The detected methanol bands on an optical depth scale. 
The dashed curve shows the $\rm 3.47~\mu m$ feature from the 
intermediate mass YSO Reipurth 50. The solid curves show the sum of laboratory spectra of methanol-rich ice mixtures at 10 K and
the  $\rm 3.47~\mu m$ feature from Reipurth 50. There is no contribution from a  $\rm 3.47~\mu m$ feature in SVS 4-5. The contents of the laboratory mixtures are indicated in the lower panel.}
\label{Methbands}
\end{figure}

A local third order polynomial was used to estimate the continuum
around the methanol feature (see Fig.~\ref{continua}). The polynomial
was fitted to the points between $\rm 3.1~\mu m$ and $\rm 3.2~\mu
m$, and the region between $\rm 3.63~\mu m$ and $\rm 3.70~\mu m$.  The
confidence in the local continuum for extraction of the methanol
profiles is high, partly due to the low order of the polynomial and
partly due to the low curvature required for the three spectra. The
uncertainty in the derived abundances is thus dominated by the main
water band continuum. The measured water ice abundances are presented in Table~\ref{Abundances}. The values were
derived by fitting a laboratory spectrum of pure amorphous water ice at 50 K to the observed
water bands and integrating the scaled laboratory spectrum from 2.5 to $\rm 3.7~\mu m$.  
None of the observed water bands are saturated. However, due to the
low S/N ratio in SVS 4-5, the fit in this source relies on the wings in the water band. The
adopted band strength is $\rm 2.0\times10^{-16}~cm~molec^{-1}$
\citep{Gerakines95}. Note that laboratory measurements of solid state band strengths may in general only
be accurate within 30\%. 

The final absorption profiles in the 3.3--$\rm 3.7~\mu m$ region are
shown in Fig.\ \ref{Methbands} on an optical depth scale. Common to
the profiles of SVS 4-9 and Cha INa 2 is a broad feature centered at $\rm 3.47~\mu m$
along with a narrow feature at $\rm 3.53~\mu m$ seen in all three sources. The former is most
likely the same feature seen in this wavelength region toward many
other young stars, generally known as the `$\rm 3.47~\mu m$ feature'
and thought to be due to diamonds or an ammonia hydrate. The 3.53 $\mu$m feature is seen toward much fewer
objects and is assigned to the $\nu_3$ CH-stretching vibration band of
solid methanol.   

To estimate the true depth of the methanol band, it is necessary to
remove the contribution from the $\rm 3.47~\mu m$ feature. A
first-order approximation is to use a high signal-to-noise profile
observed toward another YSO. We chose here the feature observed toward
the intermediate mass YSO \object{Reipurth 50} described in
\cite{ManuRE50}. The feature has a center position of $\rm 3.478~\mu m$ and a $FWHM$ of $\rm 0.12~\mu m$.
$\rm 3.47~\mu m$ features from other sources give similar results \citep{Brooke99}. The $\rm 3.47~\mu m$ band 
from Reipurth 50 was fitted to the observed spectra simultaneously with a laboratory methanol band. 

It is well-known that the $\rm 3.53~\mu m$ methanol band is sensitive
to the structure and composition of the ice \citep{Ehrenfreund99}.  Laboratory spectra are compared to
the observed methanol bands in Fig. \ref{Methbands}. A
laboratory spectrum of pure methanol does not
give a perfect fit of the exact profile to any of the observed spectra, mainly
due to the relative weakness of the secondary band of solid methanol
at $\rm 3.40~\mu m$ in the observed spectra. It is interesting to note that the
ratio of the $\rm 3.40~\mu m$ and the $\rm 3.53~\mu m$ methanol features in low-mass sources are 
similar to that of the high-mass sources of
\cite{Manu99}. A
laboratory spectrum rich in water and formaldehyde ($\rm H_2CO$) can reproduce the spectra. 
However, the H$_2$CO identification was found to be inconsistent for the massive stars \object{W
33A} and \object{RAFGL 7009S} \citep{Manu99} due to the absence of bands at 5.8
and $\rm 6.69~\mu m$. 
Other mechanisms of reducing the strength of the $\rm 3.40~\mu m$ band cannot be ruled out.
For example, irradiated pure methanol spectra annealed to temperatures higher than 60 K 
also gives good matches. For the estimate of the $\rm CH_3OH$ column densities it is assumed that
$\rm H_2CO$ does not contribute to the $\rm 3.53~\mu m$ band.

The adopted band strength of the CH-stretching mode at $\rm 3.53~\mu
m$ is $\rm 5.3\times 10^{-18}~cm~molec^{-1}$. This value is not found to
vary significantly depending on the ice mixture \citep{Kerkhof99}. The derived
abundances are summarized in Table \ref{Abundances}.

\cite{Brooke99} and \cite{Chiar96} find in general upper limits on the methanol abundance in the 
solid phase of a few \% with respect to water. We find consistent results for the rest of our observed sample, i.e. 
methanol with an abundance of more than 5\% is found toward only 3 out of a sample of $\approx 40$ low mass YSOs.

\begin{table}
\centering
\begin{flushleft}
\caption{Column densities of solid methanol and water}
\begin{tabular}{lllll}
\hline 
\hline 
Source & $\tau(\rm H_2O)$ & $N_{\rm H_2O}^a$ & $N_{\rm CH_3OH}^a$ & $N_{\rm CH_3OH}/N_{\rm H_2O}$\\
& &$10^{18}~\rm cm^{-2}$ & $10^{18}~\rm cm^{-2}$ &\\
\hline
SVS 4-5 &$4\pm1$&$7\pm2$&$1.5\pm0.1$& $0.21\pm0.05$\\
SVS 4-9 &$1.6\pm0.1$&$2.5\pm0.2$&$0.62\pm0.05$& $0.25\pm0.03$ \\
Cha INa2 &$1.5\pm0.1$&$2.2\pm0.2$&$0.3\pm0.05$& $0.14\pm0.03$\\
\hline
\end{tabular}
\begin{itemize}
\item[$^a$] Statistical $1\sigma$ errors. Uncertain laboratory estimates of band strengths may introduce a systematic error of up to 30\%.
\end{itemize}
\label{Abundances}
\end{flushleft}
\end{table}

\section{Chemical implications}

The methanol can be produced in space and in laboratory experiments
through several different pathways. Gas-phase chemistry produces only
low abundances of CH$_3$OH, of order 10$^{-9}$ with respect to H$_2$,
so that simple freeze-out of gas-phase CH$_3$OH cannot reproduce the
observed solid state abundances of $\sim$$10^{-5}$.  A possible route is
therefore thought to be grain surface hydrogenation reactions of solid
CO with atomic H, leading to H$_2$CO and eventually CH$_3$OH (Tielens
\& Hagen 1982, Charnley et al.\ 1997). Experimental results on this
reaction scheme at low temperatures are still controversial.
\cite{Watanabe} find efficient formation of $\rm H_2CO$ and $\rm
CH_3OH$ by warm ($\approx$ 300~K) hydrogen addition to CO in an $\rm
H_2O$-CO ice mixture, but the temperature of the hydrogen is not relevant for quiescent dark clouds.  
Other authors find yields which show that the
formation of methanol directly via hydrogenation of CO is inefficient
in dark clouds in the 10--25~K regime \citep{Hiraoka}. Further
experiments are needed to solve this discrepancy.

Theoretical models of grain surface chemistry indicate that the
fraction of CH$_3$OH ice with respect to H$_2$O ice, as well as the
H$_2$CO/CH$_3$OH ratio, depend on the ratio of H accretion to CO accretion on the grains and thus
on the density of the cloud (e.g., Charnley et al.\ 1997, Keane et
al.\ 2002).  In the framework of these models, solid CH$_3$OH abundances of 15--30\% with respect to
H$_2$O ice such as found here can only be produced at low densities,
$n\approx 10^4$ cm$^{-3}$.  Moreover, the region needs to be cold,
$T_{\rm dust} <20$~K, to retain sufficient CO on the grain surfaces.  
The models thus suggest that solid CH$_3$OH formation occurs primarily in the cold pre-stellar phase.
However, this fails to explain the large range of methanol abundances observed along lines of sight toward low mass YSOs in the same
star forming cloud, unless the duration of the pre-stellar phase varies strongly from source to source. As discussed by van der Tak et al.\ (2000),
comparison of solid CH$_3$OH with solid CO$_2$ can further constrain
the H/O ratio of the accreting gas and thus the density and duration
of the pre-stellar phase. For low-mass YSOs, data on solid CO$_2$ are
not yet available, but will be possible with the {\it Space Infrared
Telescope Facility} (SIRTF) via the $\rm 15.2~\mu m$ $\rm CO_2$ bending mode. 

An alternative formation route of solid CH$_3$OH may be through
ultraviolet processing \citep{WillemThesis} or proton irradiation of ices \citep{HM}. 
However, as shown by \cite{vdtak00}, the CH$_3$OH production rates in these experiments are
orders of magnitude lower than those through grain surface formation. The fact that some low-mass
sources show similarly high solid CH$_3$OH abundances as many massive sources argues further
against ultraviolet radiation playing a significant role.

It is clear that any formation scheme of solid CH$_3$OH must be able
to explain local ice mantle abundances of at least 15--30\% compared to H$_2$O ice,
regardless of the potential differences in density, temperature and
ultraviolet radiation field between high- and low-mass YSO
environments.  At the same time, such models need to explain the large
variations in solid CH$_3$OH abundances from object to object, with
abundances of less than 3--5\% found for both low- and high-mass YSOs
\citep{Chiar96, Brooke99}.  In this context, it is interesting that
abundant solid methanol is now detected toward two sources which are
members of the same dense cluster, SVS 4-5 and 4-9, while other nearby sources in 
Serpens such as CK 1 show solid methanol abundances of less that a few \% . This
suggests that the ice mantle chemistry may differ significantly
between different regions even in the same star-forming clouds or that the formation and
subsequent presence of methanol on grain surfaces is a transient
phenomenon related to strong heating of the ices. In the last scheme
the methanol may not exist for long in the solid phase before being
desorbed by the continuous heating of the ice. Finally shock processing may produce a
different ice chemistry \citep{Bergin98}. Making a survey of the
SVS 4 cluster lines of sight may help to decipher the relation of the methanol ice
to the environmental differences between `quiescent' environments as
opposed to the massive and complex star-forming environments probed so
far.

\bibliographystyle{aa} 

\begin{acknowledgements}
The authors wish to thank the VLT staff for assistance and advice, in
particular Chris Lidman for many helpful comments on the data
reduction. This research was supported by the Netherlands Organization
for Scientific Research (NWO) grant 614.041.004, the Netherlands
Research School for Astronomy (NOVA) and a NWO Spinoza grant.
\end{acknowledgements} 

\bibliography{Fd021}
\end{document}